# Fixed-complexity Vector Perturbation with Block Diagonalization for MU-MIMO Systems

Manar Mohaisen*, Bing Hui*, KyungHi Chang*, Seunghwan Ji†, and Jinsoup Joung†
* Inha University, Graduate school of IT & T, 402-751 Incheon, Korea
† Innowireless Company, 463-824 Seongnam, Korea
Email: lemanar@hotmail.com, huibing_zxo@163.com, khchang@inha.ac.kr, [shji, joung]@innowireless.co.kr

*Abstract*—Block diagonalization (BD) is an attractive technique that transforms the multi-user multiple-input multiple-output (MU-MIMO) channel into parallel single-user MIMO (SU-MIMO) channels with zero inter-user interference (IUI). In this paper, we combine the BD technique with two deterministic vector perturbation (VP) algorithms that reduce the transmit power in MU-MIMO systems with linear precoding. These techniques are the fixed-complexity sphere encoder (FSE) and the QR-decomposition with M-algorithm encoder (QRDM-E). In contrast to the conventional BD VP technique, which is based on the sphere encoder (SE), the proposed techniques have fixed complexity and a tradeoff between performance and complexity can be achieved by controlling the size of the set of candidates for the perturbation vector. Simulation results and analysis demonstrate the properness of the proposed techniques for the next generation mobile communications systems which are latency and computational complexity limited. In MU-MIMO system with 4 users each equipped with 2 receive antennas, simulation results show that the proposed BD-FSE and BD-QRDM-E outperforms the conventional BD-THP (Tomlinson Harashima precoding) by 5.5 and 7.4dB, respectively, at a target BER of $10^{-4}$.

## I. Introduction

Single-user multiple-input multiple-output (SU-MIMO) techniques, i.e., point-to-point links, have shown tremendous capacity gains without requiring additional frequency-time resources [1]. In practice, each base station (BS) communicates simultaneously with a large number of users using scarce spectral resources. Therefore, multi-user MIMO (MU-MIMO) techniques are required to enable BSs communicating with multiple users on the same frequency band and at the same time instant [2]. In analogy with the SU-MIMO case, it has been shown that when $n_T$ antennas are used to transmit to $n_U$ users with $n_R$ antennas each, the downlink sum capacity grows linearly with $\min(n_T, n_U \times n_R)$ [3].

To achieve the maximum sum capacity at the downlink of the MU-MIMO systems, several approaches relying on the information-theoretic principle of dirty-paper coding (DPC) were proposed. DPC was initially proposed by Costa, where he showed that the capacity of an interference channel with known interference is exactly the same as the interference-free channel [4]. In communication systems, the transmitted signal to one user can be seen as interference for the other users. Because this interference is known to the BS and the channel can be fed back by the users, inter-user interference (IUI) can be canceled, or highly reduced, by means of MU-MIMO precoding.

Channel inversion and regularized channel inversion are the simplest precoding techniques [5]. These approaches are sometimes referred to as zero-forcing (ZF) and minimum-mean square error (MMSE), respectively. When the channel matrix is ill-conditioned, channel inversion precoding requires high transmission power, leading to degradation in the error performance. Although regularized inversion precoding improves the conditionality of the precoding matrix, thus, reduces the required transmit power, its error performance is still mediocre.

Tomlinson-Harashima precoding (THP) limits the transmit power by introducing the non-linear modulo operation [6], [7]. As a consequence, out of constellation points at the output of the precoder are rounded off to a pre-defined range. A linearized version of the THP that consists of vector perturbation stage and IUI cancellation stage was presented in [8]. The vector perturbation stage perturb the data vector such that the transmit power is reduced. Then, the transmitted signal can be recovered at the receivers by the same modulo operation. The IUI cancellation stage can be either done successively or using any of the aforementioned linear precoders. Notice that the aforementioned precoding approaches assume that a single stream is transmitted to each user.

Block diagonalization (BD) algorithm, which transforms the MU-MIMO link into parallel SU-MIMO links, supports multi-stream transmission [9]. BD uses a precoding matrix that ensures zero IUI, where consequently users' data are processed in parallel leading to a reduction in the processing time at the BS side.

In [10], a combination of MMSE-THP and BD scheme was introduced to improve the system performance. Although THP outperforms matrix inversion precoding scheme and its regularized form, the obtained perturbation vector is not optimum. This implies that further reduction in the required transmit power can be achieved if the perturbation vector is optimized.

**Related works.** In [11], the idea of vector perturbation was introduced for single-antenna decentralized users, where the perturbation problem is solved using the sphere encoder (SE).

This work was supported by the Korea Science and Engineering Foundation (KOSEF) grant funded by the Korea government (MOST) (No. R01-2008-000-20333-0).

In [12] and [13], the vector perturbation technique is generalized for the multi-receive antenna users. This is accomplished by employing the BD algorithm.

**Contributions.** Our contributions are summarized as follows:

- We discuss the computational complexity and latency issues of the SE, and its applicability in the downlink of the multi-user multi-receive antennas MIMO systems.
- To overcome the drawbacks of the SE, we propose two deterministic BD vector perturbation techniques. These techniques are the fixed-complexity sphere encoder (FSE) [14] and the QR-decomposition with M-algorithm encoder (QRDM-E) [15] combined with the BD algorithm. Furthermore, we optimize the size of the list of candidates for the perturbation vector stage such that a tradeoff between performance and complexity is achieved.

The rest of this paper is organized as follows. In Section 2, we introduce the system model and the BD technique. In Section 3, the proposed BD vector perturbation techniques are introduces in details. Also, a review of the conventional BD vector perturbation with SE is addressed. Simulation results and discussions are introduced in Section 4 and conclusions are drawn in Section 5.

## II. SYSTEM MODEL FOR MU-MIMO WITH BLOCK DIAGONALIZATION

We consider the downlink of a MU-MIMO system, i.e., transmission from base station to users, with $n_T$ transmit antennas and $n_R$ receive antennas per user. We assume that $n_T = (n_R \times n_U)$ where $n_U$ is the number of users. Under the assumption of narrow-band flat-fading channel, the MU-MIMO channel matrix $\mathbf{H} \in \mathbb{C}^{n_R n_U \times n_T}$ is given by:

$$\mathbf{H} = \begin{bmatrix} \mathbf{H}_1^T & \mathbf{H}_2^T & \cdots & \mathbf{H}_{n_U}^T \end{bmatrix}^T, \quad (1)$$

where $\mathbf{H}_i \in \mathbb{C}^{n_R \times n_T}$ is the channel coupling the $n_T$ transmit antennas to the $n_R$ receive antennas of user $i$, and $(\cdot)^T$ denotes the matrix transpose.

### A. Block Diagonalization

The inter-user interference (IUI) can be fully canceled out using the BD algorithm. Therefore, BD transforms the MU-MIMO channel into parallel single-user MIMO (SU-MIMO) channels. The inter-symbol interference (ISI) among symbols belonging to certain user can be either removed at the transmitter by means of precoding, or at the receiver by employing spatial demultiplexing, i.e., detection. To reduce the complexity of the users' receivers, we consider that the channel effect is equalized for at the transmitter side by means of precoding.

The purpose of the BD algorithm is to find $\mathbf{B} \in \mathbb{C}^{n_T \times n_T}$ such that

$$\begin{aligned}
\mathbf{HB} &= \begin{bmatrix} \mathbf{H}_1\mathbf{B}_1 & \mathbf{H}_1\mathbf{B}_2 & \cdots & \mathbf{H}_1\mathbf{B}_{n_U} \\ \mathbf{H}_2\mathbf{B}_1 & \mathbf{H}_2\mathbf{B}_2 & \cdots & \mathbf{H}_2\mathbf{B}_{n_U} \\ \vdots & \vdots & \ddots & \vdots \\ \mathbf{H}_{n_U}\mathbf{B}_1 & \mathbf{H}_{n_U}\mathbf{B}_2 & \cdots & \mathbf{H}_{n_U}\mathbf{B}_{n_U} \end{bmatrix} \\
&= \begin{bmatrix} \mathbf{H}_{\text{eff},1} & \mathbf{0}_{n_R} & \cdots & \mathbf{0}_{n_R} \\ \mathbf{0}_{n_R} & \mathbf{H}_{\text{eff},2} & \cdots & \mathbf{0}_{n_R} \\ \vdots & \vdots & \ddots & \vdots \\ \mathbf{0}_{n_R} & \mathbf{0}_{n_R} & \cdots & \mathbf{H}_{\text{eff},n_U} \end{bmatrix},
\end{aligned} \quad (2)$$

where $\mathbf{0}_{n_R}$ is the $n_R \times n_R$ zero matrix and $\mathbf{H}_{\text{eff},i} = \mathbf{H}_i\mathbf{B}_i$ is the effective channel matrix of user $i$ after the BD. To this end, we define the matrix

$$\mathbf{H}_{\bar{i}} = \begin{bmatrix} \mathbf{H}_1^T & \cdots & \mathbf{H}_{i-1}^T & \mathbf{H}_{i+1}^T & \cdots & \mathbf{H}_{n_U}^T \end{bmatrix}^T, \quad (3)$$

which is obtained by simply removing the channel matrix of user $i$ from the system channel matrix $\mathbf{H}$. The singular value decomposition (SVD) of $\mathbf{H}_{\bar{i}}$ is formed as follows:

$$\mathbf{H}_{\bar{i}} = \mathbf{U}_{\bar{i}} \Sigma_{\bar{i}} \begin{bmatrix} \mathbf{V}_{\bar{i}}^{(1)} & \mathbf{V}_{\bar{i}}^{(0)} \end{bmatrix}^H, \quad (4)$$

where the columns of $\mathbf{V}_{\bar{i}}^{(0)}$ are the right singular vectors corresponding to the zero singular values of $\mathbf{H}_{\bar{i}}$. Since the columns of $\mathbf{V}_{\bar{i}}^{(0)}$ lead to zero IUI, they will be potential beamformers for user $i$. Therefore, a linear combination of these vectors is found to form the beamforming matrix $\mathbf{B}_i$. To accomplish this, the SVD of $\mathbf{H}_i \mathbf{V}_{\bar{i}}^{(0)}$ is formed as follows:

$$\mathbf{H}_i \mathbf{V}_{\bar{i}}^{(0)} = \mathbf{U}_i \Sigma_i \begin{bmatrix} \mathbf{V}_i^{(1)} \end{bmatrix}^H, \quad (5)$$

where $\mathbf{H}_i \mathbf{V}_{\bar{i}}^{(0)}$ is considered to be full-rank. $\mathbf{B}_i$ is then equal to $\mathbf{V}_{\bar{i}}^{(0)} \mathbf{V}_i^{(1)}$, and the transmit beamforming matrix $\mathbf{B} \in \mathbb{C}^{n_T \times n_T}$ is given by:

$$\mathbf{B} = \begin{bmatrix} \mathbf{B}_1 & \mathbf{B}_2 & \cdots & \mathbf{B}_{n_U} \end{bmatrix}. \quad (6)$$

Notice that when the number of users increases for a fixed $n_T$, the degrees of freedom at the base station are spent in the IUI nulling process, leading to reduction in the transmit diversity of the array.

## III. THE PROPOSED BLOCK DIAGONALIZATION VECTOR PERTURBATION

Applying the BD algorithm to $\mathbf{H}$ transforms the MU-MIMO channel into $n_U$ parallel SU-MIMO channels with zero IUI. Therefore, user streams can be processed in parallel, leading to the reduction in the precoding latency. Fig. 1 shows the resulting end-to-end SU-MIMO system for user $i$. The details of Fig. 1 are provided in the following Subsections. In this paper, all users' symbols are withdrawn from the same constellation set. Hence, we did note employ any kind of power loading techniques, and the beamforming matrix $\mathbf{F}_i$ is equivalent to $\mathbf{B}_i$. In the sequel, we consider the data processing for user $i$, which is identical to those of other users. Also, we consider that the SU-MIMO system for user $i$ is mapped to the $N$-dimensional real Euclidean space for $N = 2n_R$, where real vectors and matrices are represented by italic boldface symbols.

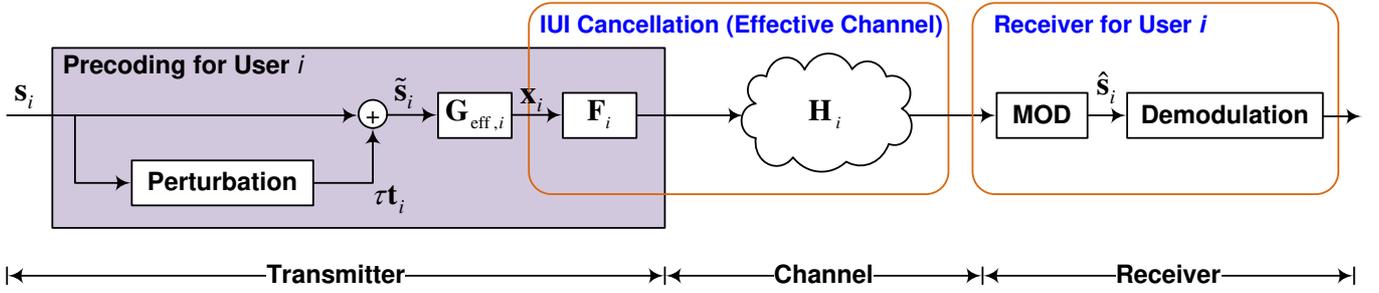

Fig. 1. Structure of the broadcast MIMO precoding system using block diagonalization and vector perturbation techniques for user $i$.

### A. Vector Perturbation for MU-MIMO Systems and a Review of the Sphere Encoder

The goal of the vector perturbation techniques is to generate the vector $\tilde{s}_i$ from the data vector $s_i$, such that the norm of $G_{\text{eff},i}\tilde{s}_i$ becomes smaller than that of $G_{\text{eff},i}s_i$. Here, $G_{\text{eff},i}$ is the precoding matrix for user $i$. The precoding matrix $H_{\text{eff},i}$ equals $H_{\text{eff},i}^{-1}$ and $(H_{\text{eff},i}H_{\text{eff},i}^H + \alpha I)^{-1}H_{\text{eff},i}^H$ for the linear ZF and MMSE precoders, respectively. The regularization factor $\alpha$ equals $(N\sigma_n^2/P_i)$, where $\sigma_n^2$ and $P_i$ are the single-sided noise variance and the total transmit power for user $i$. In other words, the vector perturbation technique is employed to reduce the required transmission power. The perturbed vector $\tilde{s}_i$ is then derived from the THP technique as follows:

$$\tilde{s}_i = s_i + \tau t, \qquad (7)$$

where $\tau$ is an integer that depends on the used modulation scheme, and $t$ is an $N$-dimensional integer vector. In [11], $\tau$ is given by:

$$\tau = 2\left(|c_{max}| + \Delta/2\right), \qquad (8)$$

where $|c_{max}|$ is the absolute value of the symbol with the largest magnitude, and $\Delta$ is the spacing between any two neighbor symbols. Since the precoding equalizes for the channel effect, the received vector at user $i$ is given by:

$$y_i = \tilde{s}_i + n_i. \qquad (9)$$

where $n_i$ is the additive-white Gaussian noise vector with covariance matrix $\sigma_n^2 I$. At the receiver, the original data vector $s_i$ is recovered, without knowledge of vector $t$, using the non-linear modulo operation as follows:

$$\hat{s}_i = \text{mod}(y_i), \qquad (10)$$

where $\text{mod}(\cdot)$ is the modulo operation that reduces the range of the received signal to the interval $[-K, K]$, where $K$ depends on the used modulation scheme [18]. Specifically, $K = \sqrt{|\Omega|}$, where $|\Omega|$ is the cardinality of the modulation set $\Omega$. For instance, $K = 2$ and $4$ for QPSK and 16-QAM modulation schemes, respectively.

The vector $t$, introduced in (7), is found by solving the following $N$-dimensional integer lattice problem:

$$\begin{aligned} t &= \arg\min_{t \in \mathbb{Z}^N} \left\{ (s + \tau t)^H G_{\text{eff},i}^H G_{\text{eff},i}(s + \tau t) \right\}, \\ &= \arg\min_{t \in \mathbb{Z}^N} \|G_{\text{eff},i}(s + \tau t)\|^2. \end{aligned} \qquad (11)$$

In [12] and [13], authors outlined the benefit of employing the BD algorithm in transforming the $(N \times n_U)$-dimensional lattice search problem into $n_U$ parallel $N$-dimensional search problems. Clearly, this leads to the reduction in the precoding latency. However, we state the following remarks about using the SE to solve the lattice problem in (11):

1) **Sphere encoder:** Authors used the sphere encoder (SE) which was originally introduced by Hochwald *et al.* in [11]. Although the SE achieves great reduction in the required transmission power, it suffers from the following drawbacks:
    - **Worst-case complexity:** Although the average complexity of the SE is polynomial in the problem size [16], its worst-case complexity is exponential, i.e., comparable to that of the *brute-force search*[1]. Therefore, in computational complexity limited communication systems, the SE becomes inapplicable.
    - **Maximum latency:** Because SE is sequential in the tree search phase, this limits the possibility for efficient hardware implementation by pipelining.
    - **User-dependent precoding latency:** Because the computational complexity of the SE depends, among other factors, on the conditioning of $H_{\text{eff},i}$, the precoding latency may differ from a user to another. Hence, the vector-perturbation stage for a user with ill-conditioned effective channel matrix is more time consuming than that for a user with well-conditioned channel matrix. Therefore, the processing latency at the transmitter for precoding a data vector is equivalent to the maximum latency to precode $s_i$, for $i = 1, 2 \cdots, n_U$. Thus, additional latency overhead is introduced at the transmitter side due to the random complexity of the vector perturbation stage using the SE.

[1]In fact, Jalden and Ottersten have shown in [17] that even the average computational complexity of the sphere decoder is exponential in the problem size for a fixed signal to noise ratio.

2) **Choice of the set of candidates for $t$:** Because the elements of $t$ can have any integer value, the set of candidates for $t$ should be a truncated subset of $\mathbb{Z}$ to restrict both the precoding latency and computational complexity. Authors in [11], [12], and [13] did not impose any restriction on the size of the set of candidates, leading to a huge worst-case complexity.

### B. Problem Formulation

To solve (11) successively in the case of the ZF precoding, the LQ factorization of the effective channel matrix $\boldsymbol{H}_{\text{eff},i}$ is required. Let the transpose of $\boldsymbol{H}_{\text{eff},i}$ be factorized into the product of a unitary matrix $\boldsymbol{Q}$ and an upper triangular matrix $\boldsymbol{R}$, then, the search problem in (11) can be simplified to:

$$\boldsymbol{t} = \arg\min_{\boldsymbol{t}\in\mathbb{Z}^N} \|\boldsymbol{L}(\boldsymbol{s}_i + \tau\boldsymbol{t})\|^2,$$

$$= \arg\min_{\boldsymbol{t}\in\mathbb{Z}^N} \sum_{n=1}^{N} \left\| L_{n,n}(s_n + \tau t_k) + \sum_{j=1}^{n-1} L_{n,j}(s_j + \tau \hat{t}_j) \right\|^2, \quad (12)$$

where $s_n$ is the $n$-th element of $\boldsymbol{s}_i$, and the lower triangular matrix $\boldsymbol{L}$ equals $(\boldsymbol{R}^{-1})^T$. Also, $\hat{t}_k$ and $t_k$ represent the retained candidate and a possible candidate for $t$, respectively, at the $k$-th precoding level. In the case of the MMSE precoder, the extended matrix $\tilde{\boldsymbol{H}}_{\text{eff},i} = [\boldsymbol{H}_{\text{eff},i}^T \ \sqrt{\alpha}\boldsymbol{I}]^T$ is factorized, where $\boldsymbol{L}$ again equals $(\boldsymbol{R}^{-1})^T$. Due to the QR property

$$\begin{bmatrix} \boldsymbol{H}_{\text{eff},i}^T \\ \sqrt{\alpha}\boldsymbol{I} \end{bmatrix} = \begin{bmatrix} \boldsymbol{Q}_1 \\ \boldsymbol{Q}_2 \end{bmatrix}\boldsymbol{R} = \begin{bmatrix} \boldsymbol{Q}_1\boldsymbol{R} \\ \boldsymbol{Q}_2\boldsymbol{R} \end{bmatrix}, \quad (13)$$

it holds that $\boldsymbol{R}^{-1} = \boldsymbol{Q}_2/\sqrt{\alpha}$ [19]. By definition $\sqrt{\alpha}$ is a strictly positive real number, then it does not affect the search result in (12). Therefore, $\boldsymbol{L} = \boldsymbol{Q}_2^T$ also leads to the required perturbation without the need for explicitly inverting $\boldsymbol{R}$.

The elements of $\boldsymbol{t}$ are drawn from the symmetric integer set

$$\mathcal{A} = [-a, -a+1, \cdots, a-1, a], \quad (14)$$

where $a$ is a positive integer chosen to achieve a tradeoff between performance and complexity of the vector-perturbation stage. Hereafter, $T = (2a+1)$ denotes the number of elements of the set $\mathcal{A}$.

In the following Subsections, we introduce the proposed BD vector perturbation techniques that overcome the aforementioned drawbacks of the BD-SE with a tolerable sub-optimality in solving the integer lattice problem in (11).

### C. Fixed-complexity Sphere Encoder (FSE)

The tree-search phase of the FSE algorithm consists of the following two steps:
- **Full expansion:** At the first $p$ tree search levels, the retained branches are expanded to all possible nodes, and all the resulting branches are retained for the next level.
- **Single expansion:** All retained branches in the precedent level are independently expanded to all possible nodes. Then, the accumulative metrics of the resulting branches are calculated using (12), and only the branch with the smallest accumulative metric is retained for the next level.

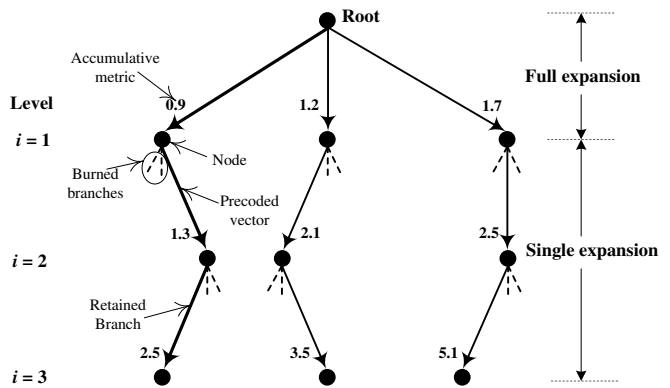

Fig. 2. Example of the FSE for $T = N = 3$.

At the last search level, the metrics of the obtained perturbed vectors $\tilde{\boldsymbol{s}}_1, \tilde{\boldsymbol{s}}_2, \cdots, \tilde{\boldsymbol{s}}_T$ are compared, and the vector that has the smallest metric is precoded and transmitted.

Fig. 2 depicts an example of the FSE for $T = N = 3$ and $p = 1$. At the first level, i.e., $i = 1$, the root node is expanded to all possible combinations $(s_1 + \tau t_k)$ for $t_k \in \{-1, 0, 1\}$. The metrics of the resulting branches are calculated via (12), and all the branches are retained for the next search level. Each retained branch at level 1 is expanded to the three possible combinations $(s_2 + \tau t_k)$ for $t_k \in \{-1, 0, 1\}$, and the branch with the smallest accumulative metric is retained. This strategy is repeated at level 3, where the leaf, i.e., perturbed vector, that has the lowest accumulative metric is precoded and transmitted.

The advantages of using the BD-FSE over the conventional BD-SE are summarized as following:
- BD-FSE algorithm has a fixed complexity that is independent of the channel conditionality. A tradeoff between complexity and performance is achieved by selecting an appropriate value for $a$. Therefore, for an equal number of receive antennas per user, the latency of the vector perturbation stage is the same for all the users. This avoids the time wasting problem of the conventional BD-SE, resulting when the latency of the vector perturbation stage varies from a user to another.
- The vector perturbation stage of the BD-FSE is parallel, as shown in Fig. 2, leading to efficient hardware implementation by pipelining. This reduces the precoding latency which is an important issue in the communication systems of beyond third generation (3G) [20].

For $(p = 1)$, the proposed FSE visits only $(N \times T)$ nodes to obtain the perturbed vector with a computational complexity much lower than that of the QRDM-E as will be shown in the next Section.

### D. QR-decomposition with M-algorithm encoder (QRDM-E)

In QRDM-E, the best $M$ branches that have the least accumulative metrics are retained at each encoding level. To accomplish a fair comparison with the FSE, $M$ is set to $T$. Therefore, at the first tree-search stage, the best $M$ branches are retained for level 2. At level 2, the retained branches are expanded to all possible combinations of $(s_2 + \tau t_k)$. The resulting $M^2$ branches are sorted according to their accumulative metrics calculated via (12), where only the $M$ branches with the smallest accumulative metrics are retained for level 3. This strategy is repeated up to the last encoding level, where the perturbed vector $\tilde{s}_i$ that has the smallest accumulative metric is precoded and transmitted.

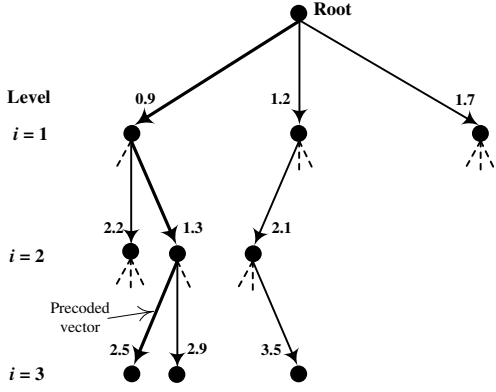

Fig. 3. Example of the QRDM-E for $T = N = 3$.

Fig. 3 depicts an example of the QRDM-E for $T = N = M = 3$. At each encoding level, only the best three branches with the least accumulative metrics are retained. In contrast to the SE which visits $(\sum_{i=1}^{N} M^i)$ nodes as the worst-case complexity, QRDM-E has a fixed complexity where it only visits $(M + (N-1)M^2)$ nodes.

## IV. SIMULATION RESULTS AND DISCUSSIONS

In this Section we optimize the size of the set of candidates $\mathcal{A}$. Then, we evaluate the bit error rate (BER) performance of the proposed BD vector perturbation techniques in an $(n_T, n_U, n_R)$ MU-MIMO systems, with $n_T = n_U \times n_R$. The conventional THP approach is considered as the special case of the vector perturbation techniques when only the branch that has the least accumulative metric is retained at each tree search level. Also, due to its superior performance compared to the ZF criterion, the MMSE criterion is used to construct the precoding matrix $\bm{G}_{eff}$.

Fig. 4 depicts the BER performance of the proposed BD vector perturbation techniques in $(8, 2, 4)$ and $(8, 4, 2)$ MU-MIMO systems at SNR of 20 and 25dB, respectively, for several values of $T$ and using 4-QAM. We remark that for both BD-FSE and BD-QRDM-E techniques, a maximum improvement is achieved when moving from $(T = 3)$ to $(T = 5)$. In the case of the BD-FSE, additional improvement in the BER performance can't be achieved for $T > 7$. On the other hand, small additional improvement is achieved in the case of the BD-QRDM-E algorithm for $T > 7$. As a tradeoff between performance and complexity, we use $T = 7$ in the sequel; that is,

$$\mathcal{A} = \{-3, -2, \cdots, 2, 3\}. \tag{15}$$

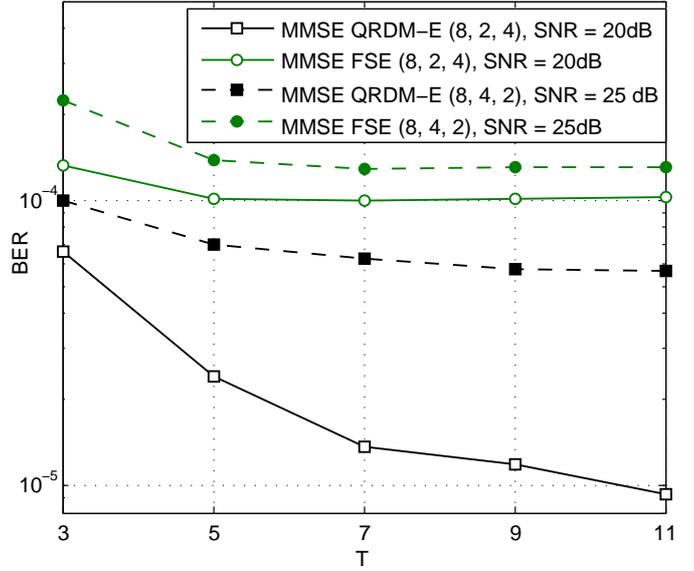

Fig. 4. Choice of the number of elements in the set of candidates $\mathcal{A}$ using 4-QAM.

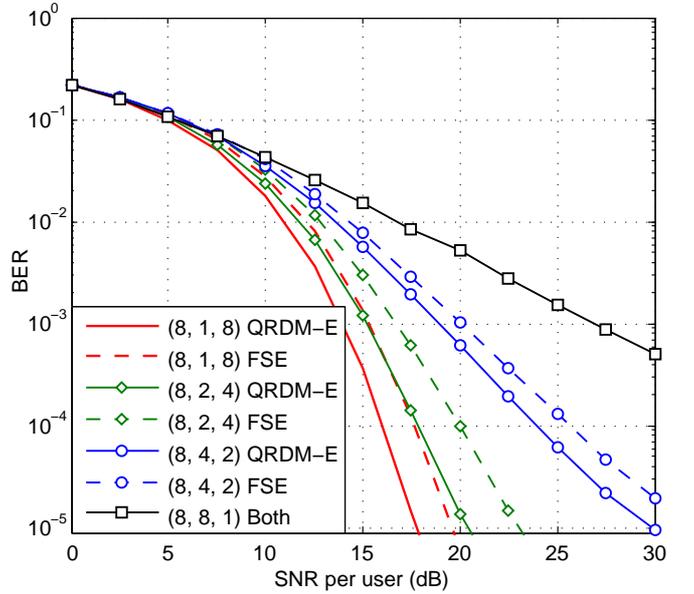

Fig. 5. BER performance of the proposed BD vector perturbation techniques for $T = 7$, and using 4-QAM modulation.

Fig. 5 depicts the BER performance of the proposed BD vector perturbations techniques for several system configurations. DB-QRDM-E outperforms BD-FSE for all system configurations. For instance, at target BER of $10^{-4}$, BD-QRDM-E outperforms BD-FSE by 1.65, 2.1, and 1.7dB in $(8, 1, 8)$, $(8, 2, 4)$, and $(8, 4, 2)$ MU-MIMO systems, respectively. The advantage of the BD-FSE compared to the BD-QRDM-E is that it has a parallel tree-search, leading to reduction in the latency of the vector perturbation stage.

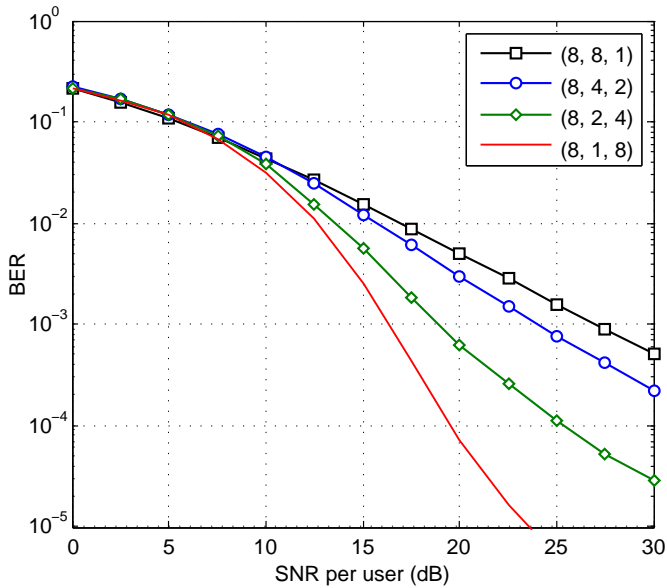

Fig. 6. BER performance of the conventional MMSE-THP with BD for $T = 7$, and using 4-QAM modulation.

Fig. 6 shows the BER performance of the introduced BD-THP algorithm. At BER of $10^{-4}$, BD-THP lags the performance of BD-QRDM-E and BD-FSE algorithms by 7.4 and 5.2dB, respectively, in $(8, 2, 4)$ MU-MIMO system. This degradation is due to the non-optimality of the obtained perturbation vector. Furthermore, we remark that a floor in the BER performance of BD-THP scheme appears at high SNR values, which is also due to the non-optimality of the vector perturbation stage.

At high SNR values, the slope of the BER curves is directly proportional to the achieved diversity order. From Fig. 5 and 6, we remark that the achieved diversity orders by the proposed techniques are equivalent and linearly proportional to the number of receive antennas per user. In contrast, the diversity order attained by the conventional BD-THP technique tend to be unity, despite that an improvement in the BER is achieved when the number of receive antennas per user is increased for a fixed $n_T$.

## V. Conclusions

In this paper, we proposed the combination of fixed-complexity FSE and QRDM-E multiuser vector perturbation techniques with the block diagonalization algorithm. The block diagonalization transforms the MU-MIMO channel into parallel SU-MIMO channels with zero inter-user interference. FSE or QRDM-E technique is used in the vector perturbation stage that aims to reduce the transmission power. In the proposed algorithms, a tradeoff between computational complexity and performance is achieved by controlling the size of the set of candidates at the vector perturbation stage. Using extensive simulations, the optimum size of the set of candidates is obtained. Also, due to its parallel tree-search stage, FSE can be pipelined, leading to tremendous reduction in the precoding latency. Therefore, the proposed algorithms are implementation efficient as compared with the conventional BD-SE technique, which has a random complexity and sequential tree-search stage. In terms of BER performance, the proposed techniques outperform the conventional BD-THP technique by more than 5dB in $(8, 2, 4)$ MU-MIMO system. Therefore, due to their low and fixed complexities, the proposed algorithms are strong candidates for implementation in the future communication systems.